\documentclass[useAMS,usenatbib]{mn2e}
\usepackage{natbib}
\usepackage{color,graphicx} % graphicx needed for \includegraphics
\usepackage{amsmath}        % need ams stuff    %
\usepackage{amsfonts}       % to make equations %
\usepackage{amssymb}        % look nice         %
\usepackage{wasysym}        % wasysym needed for certain symbols

\begin{document}
\author[K.~Liu et al.]{K.~Liu,$^{1,2}$ E.~F.~Keane,$^{1,2}$
  K.~J.~Lee,$^{2}$ M.~Kramer,$^{1,2}$ J.~M.~Cordes$^{3}$ \newauthor \& M.~B.~Purver$^{1}$\\
  $^{1}$University of Manchester, Jodrell Bank Centre of Astrophysics, Alan-Turing
  Building, Manchester M13 9PL, UK\\
  $^{2}$Max-Planck-Institut f\"{u}r
  Radioastronomie, Auf dem H\"{u}gel 69, D-53121 Bonn, Germany \\
  $^{3}$Astronomy Department, Cornell University, Ithaca, NY 14853,
  USA}

\title[Profile stability and pulse jitter]{Profile shape stability and phase jitter analyses of millisecond pulsars}

\maketitle

\begin{abstract}
  Millisecond pulsars (MSPs) have been studied in detail since their
  discovery in 1982. The integrated pulse profiles of MSPs appear
  to be stable, which enables precision monitoring of the pulse
  times of arrival (TOAs). However, for individual pulses the shape and arrival
  phase can vary dramatically, which is known as pulse jitter.
  In this paper, we investigate the stability of integrated pulse profiles for 5 MSPs,
  and estimate the amount of jitter for PSR~J0437$-$4715. We do not
  detect intrinsic profile shape variation based on integration
  times from $\sim10$ to $\sim100$\,s with the provided instrumental sensitivity.
  For PSR~J0437$-$4715 we calculate the jitter parameter to be $f_J=0.067\pm0.002$,
  and demonstrate that the result is not significantly affected by instrumental
  TOA uncertainties. Jitter noise is also found to be independent of
  observing frequency and bandwidth around 1.4\,GHz on frequency
  scales of $<100$\,MHz, which supports the idea that pulses within
  narrow frequency scale are equally jittered. In addition, we point out that
  pulse jitter would limit TOA calculation for the
  timing observations with future telescopes like the Square Kilometre
  Array and the Five hundred metre Aperture Spherical Telescope.
  A quantitative understanding of pulse profile stability and
  the contribution of jitter would enable improved TOA calculations,
  which are essential for the ongoing endeavours in pulsar timing, such as the
  detection of the stochastic gravitational wave background.
\end{abstract}


\section{Introduction}\label{sec:intro}
Millisecond pulsars (MSPs) have been shown to exhibit highly stable
rotational behaviour \citep[e.g.][]{vbc+09}. They are a vital tool
in investigating the physical environment of the pulsars, via
precise monitoring of pulse times of arrival (TOAs), the so-called
pulsar timing technique. Previous work using precision pulsar timing
has already yielded tests of General Relativity in the strong field
regime \citep[e.g.][]{tw89,ksm+06}, improvements of neutron star
equation of state models \citep[e.g.][]{lp07,dpr+10,opr+10}, and
studies of the interstellar medium (ISM) structure
\citep[e.g.][]{yhc+07}. Via pulsar timing arrays (PTAs), upper
bounds have already been placed on the stochastic gravitational wave
background \citep{jhv+06,hlj+11,ych+11}. The next generation of
radio telescopes (such as the Five hundred metre Aperture Spherical
Telescope, FAST and the Square Kilometre Array, SKA) will
significantly improve upon the available instrumental sensitivity
allowing timing to much higher precision and for many more sources.
This tremendous advance in hardware will significantly increase the
sensitivity of future PTAs and allow more detailed studies of the
gravitational wave background, such as its polarisation and speed of
propagation \citep{ljp08,ljp+10}. Individual gravitational wave
sources, such as supermassive black hole binaries at the centre of
galaxies can also be identified via a future PTA study
\citep[e.g.][]{lwk+11,srr+11}.

The extreme timing precision required to reach the aforementioned
scientific goals is more readily achievable with MSPs because of
their short spin periods, their regular rotational behaviour and,
last but not least, their greatly stable average pulse shapes. In
general, single pulses from a pulsar show significant shape
modulation. Up to now, several types of variable behaviour have been
observed within the population of pulsars. These include intrinsic
pulse-to-pulse changes caused by random phase jitter of individual
pulses \citep{cd85,cor93}, systematic position changes of
sub-components named ``sub-pulse drifting''
\citep[e.g.][]{dc68,cor75,es03}, switching between two or more
profile shapes on both short and long timescales known as
``mode-changing'' \citep[e.g.][]{crb78,fbw+81,lhk+10}, and state
changes of intrinsic flux density mentioned as ``nulling''
\citep[e.g.][]{bac70,wmj07}. Studies of bright individual pulses
from MSPs have shown only the first type of pulse-to-pulse variation
\citep{cstt96,jak+98}, which can cause fluctuation in the phase of
integrated profiles and introduce TOA uncertainty in addition to the
radiometer noise \citep{cs10}. This is the effect mainly addressed
within the framework of this paper.

Statistically, as the single pulses are clearly unstable, shape
differences are expected to exist between an integrated profile over
a short integration time and a standard template formed by averaging
many more pulses. Note that the shape difference, if sufficiently
large, would influence the accuracy of TOA estimation by the
standard cross-correlation approach \citep{lvk+11}. Consequently, it
is important to evaluate whether or not the shape mismatch is
significant compared with the system noise level. There have already
been studies investigating the shape correlation between integrated
profiles (or even single pulses) and a pre-formed template
\citep{hmt75,rr95,jak+98,jap01}. The results show clear shape
difference for young pulsars and a less significant or even
undetectable difference for MSPs with the given sensitivity.

If the TOA uncertainties can be shown to be mostly due to radiometer
noise and phase jitter, based on an assumed model, the amount of
jitter can be estimated from timing on a short timescale and such
analysis has already been carried out for PSR~J1713+0747
\citep{cs10}. The result can be used both for the error analysis in
timing and as an input for jittered shape correction approaches
\citep{mldr11,ovh+11}.

The structure of this paper is as follows. In
Section~\ref{sec:theory} we introduce the approach used for
evaluating profile stability and jitter estimation. The
observations, instruments and data pre-processing techniques used
are described in Section~\ref{sec:obs}. In Section~\ref{sec:result}
we present the results before drawing our conclusions in
Section~\ref{sec:conclu}.

\section{Profile shape and pulse jitter analysis} \label{sec:theory}
\subsection{Stability Analysis}\label{ssec:stab theory}
The shape and phase instability of single pulses will induce shape
modulation for integrated profiles, which can be mitigated by
increasing the number of pulses added. The similarity between an
observed integrated profile $p$ and a normalised standard shape $s$,
obtained from previous observations, is simply evaluated by the
correlation coefficient $\rho$, which is defined as:
\begin{equation}\label{eq:ccdef}
  \rho=\frac{\displaystyle\sum_{i}(s_{i}-\bar{s})(p_{i}-\bar{p})}{\displaystyle\sqrt{\sum_{i}(s_{i}-\bar{s})^{2}}\sqrt{\sum_{i}(p_{i}-\bar{p})^{2}}},
\end{equation}
where $i$ stands for the sample number of the data points across the
profile. It is shown in Appendix~\ref{app:scaling&err&rn} that,
assuming an identical intrinsic shape for the observed profile and
standard, there is a scaling law between $\rho$ and the profile peak
signal-to-noise ratio (SNR, defined as pulse peak amplitude divided
by root-mean-square of the noise) of: $(1-\rho)\propto\rm SNR^{-2}$.
Note that this power law is followed only in the high SNR regime
(e.g. for SNR$\gtrsim20$). From the scaling law, we define a shape
constant, $\mathfrak{C}$, related only to the intrinsic profile
shape as:
\begin{equation}
\mathfrak{C}\equiv {\rm
SNR}\cdot\sqrt{1-\rho}\simeq\left(\frac{n_{\rm
samp}}{\displaystyle2\sum_{i}(s_{i}-\bar{s})^2}\right)^{1/2},\label{eq:SCdef}
\end{equation}
where $n_{\rm samp}$ is the number of time samples of a profile and
$i=1,2,\cdots n_{\rm samp}$. Once the SNR and $\rho$ are measured,
respectively, the shape constant can then be determined and used to
compare with the value calculated directly from the waveform of the
standard.

If the profile integration increases the SNR as expected as $\rm
SNR\propto\sqrt{N}$, where $N$ is number of pulses folded, then we
reach the relation between $N$ and $\rho$ as: $(1-\rho)\propto
N^{-1}$. However, the scaling is not obeyed if the SNR varies
significantly from pulse to pulse, which can be caused by intrinsic
flux variation, system temperature changes and diffractive
scintillation. So in our analysis we use the effective pulse number
$N_{\rm efc}$, the number of pulses weighted by SNR to account for
the variation of the profile SNR which does indeed scales as $\rm
SNR\propto \sqrt{N_{\rm efc}}$ and hence $1-\rho\propto N^{-1}_{\rm
efc}$ \citep{lvk+11}.

\subsection{Phase Jitter}\label{ssec:jitter}
Single pulse instability can also cause TOA fluctuations of
integrated profiles. This phase jitter could, in principle, be
investigated directly from single pulse data, although the
sensitivity achieved by the current instruments may not enable
sufficient SNR for carrying out the study on all pulses within a
narrow band. Another approach is to perform timing using integrated
profiles on short timescales and to estimate the amount of jitter
from the timing residuals. As a first-order approximation, by
assuming an identical shape for single pulses and a
Gaussian-distributed central phase probability, the contribution of
pulse jitter to the uncertainty of TOA can be calculated as in
\cite{cs10}. In brief, the measurement error of TOAs on short
timescales (e.g. several hours) can be summarised as:
\begin{equation}
\sigma_{\rm total}^{2}=\sigma_{\rm rn}^{2}+\sigma_{\rm
J}^{2}+\sigma_{\rm scint}^{2}+\sigma^2_{0}, \label{eq:sigmas}
\end{equation}
where $\sigma_{\rm rn}$, $\sigma_{\rm J}$, $\sigma_{\rm scint}$ and
$\sigma_{0}$ correspond to uncertainty induced by radiometer noise,
pulse jitter, instability of short-term diffractive scintillation,
and all other possible contributions (faults in timing model,
instrumental digitisation artefacts, polarisation calibration error,
etc), respectively. Following \cite{dr83}, \cite{cwd+90} and
\cite{cs10}, we have
\begin{eqnarray}
\sigma_{\rm rn}^{2}&=&\frac{\Delta}{N\times SNR_{1}^{2}\int[U'(t)]^{2}dt}, \label{eq:sigma_rn}\\
\sigma_{\rm J}^{2}&=&f_J^{2}\frac{\int U(t)t^{2}dt}{N\int U(t)dt},
\label{eq:sigma_j} \\
\sigma_{\rm scint}^{2}&=&\frac{t^2_{\rm d}}{N_{\rm
scint}}.\label{eq:sigma_scint}
\end{eqnarray}
Here $N$ is the number of pulses, $SNR_{1}$ is the signal-to-noise
ratio for a single pulse, $U(t)$ is the profile waveform, $\Delta$
is the sampling time, $f_{J}$ is the width of the Gaussian
probability of single pulse phase in units of the pulse width,
$t_{\rm d}$ is the pulse broadening timescale, and $N_{\rm scint}$
is the number of scintles contained by the integration. We can see
that the ratio of $\sigma_{\rm rn}$ to $\sigma_{\rm J}$ is
proportional to the equivalent single pulse SNR. The value of
$SNR_{1}$ for MSPs, based on current observations, is mostly far
less than unity, leading to the case where the white noise term is
dominant in TOA uncertainty. However, the contribution by phase
jitter will become more significant when timing is carried out by
the next generation of radio telescopes, which will have a
significant improvement in sensitivity of $1-2$ orders of magnitude
over current systems \citep{nan06,sac+07,lvk+11}. In this case, TOA
error prediction based solely on radiometer noise will be incorrect.

The statistics of the timing residuals which are obtained by
subtracting a timing model from the measured TOAs, can be evaluated
via a reduced $\chi^{2}$ value given by
\begin{equation}
\chi^2_{\rm
rec}=\frac{1}{N-n-1}\chi^2=\frac{1}{N-n-1}\sum_i\left(\frac{\Delta
x_i}{\sigma_i}\right)^2,
\end{equation}
where $N$ is the number of residuals, $n$ is the number of fitted
parameters, and $\Delta x_i$, $\sigma_i$ are the $i$th residual and
its corresponding measurement error, respectively. Normally,
$\sigma_i$ accounts for only the uncertainty of radiometer noise
$\sigma_{\rm rn}$ in Eq.~(\ref{eq:sigma_rn}), which can be obtained
from the template matching technique \citep{tay92}. Theoretically,
if the timing residuals are dominated by radiometer noise, a timing
solution with residuals of $\chi^{2}_{\rm rec}\simeq1$ will be
expected. The existence of other types of noise would make the
$\sigma_i$ underestimated and deviate $\chi^{2}_{\rm rec}$ from
unity. The contribution from the diffractive scintillation can be
estimated by Eq.~(\ref{eq:sigma_scint}). Here $t_{\rm d}$ can be
obtained from the {\it NE2001 Galactic Free electron Density Model}
\citep{cl02}, and the number of scintles is assessable via either a
dynamic spectrum or more detailed calculations in \cite{cwd+90}. If
the additional noise is mostly from pulse jitter, one can estimate
the jitter noise $\sigma_J$ by adding its contribution into
$\sigma_i$ to have $\chi^{2}_{\rm rec}=1$. Accordingly, the value of
the jitter parameter $f_J$ can be derived based on
Eq.~(\ref{eq:sigma_j}), and the probability density distribution of
$f_J$ can be calculated from the standard $\chi^{2}$ distribution,
given by
\begin{equation}
f(x;k)=\frac{1}{2^{k/2}\Gamma(k/2)}x^{k/2-1}e^{-x/2},
\end{equation}
where $x$ is the $\chi^{2}$ value and $k=N-n-1$ is the degrees of
freedom.

\section{Observations}\label{sec:obs}
\begin{table}
\centering \caption{Details of observations used in this paper. The
symbols $t_0$, $N_0$ and $T_{\rm obs}$ represent the shortest
integration time in the dataset, the number of time dumps and the
observational length, respectively. The asterisk (*) denotes that
the data were pre-processed by the DFB and the others are from
CPSR2.}\label{tab:data}
\begin{tabular}[c]{cccccc} \\
\hline
Pulsar Name &MJD    &Receiver &$t_0$ (s) &$N_0$    &$T_{\rm obs}$ (hr)\\
\hline
J0437$-$4715  &53576  &MB       &16.7772   &1595            &8.7\\
             &53621  &MB       &16.7772   &1498            &8.9\\
             &53864  &MB       &67.1088   &261             &6.9\\
             &~$54095^{*}$  &H-OH     &60.0017   &125             &2.4\\
             &54222  &H-OH     &67.1088   &135             &9.8\\
             &54226  &H-OH     &67.1088   &152             &8.9\\
J1022+1001    &53260  &MB       &16.7772   &114             &0.5\\
J1603$-$7202  &53166  &H-OH     &16.8099   &114             &0.5\\
J1713+0747    &53221  &H-OH     &16.7771   &215             &3.3\\
J1730$-$2304  &53145  &H-OH     &16.7772   &114             &0.5\\
\hline
\end{tabular}
\end{table}
Frequent observations, typically weekly, of more than 20 MSPs are
performed at the Parkes 64-m radio telescope \citep{vbc+09}. Here we
use the data from five sources (PSR~J0437$-$4715, PSR~J1022+1001,
PSR~J1603$-$7202, PSR~J1713+0747, PSR~J1730$-$2304), collected using
either the Parkes 20-cm multibeam (MB) receiver \citep{smt+96} or
the `H-OH' receiver. The data were processed online with the
Caltech-Parkes-Swinburne Recorder 2 (CPSR2), a 2-bit coherent
de-disperser back-end that records two 64-MHz wide observing bands
simultaneously \citep{hbo06}. These bands were centred at observing
frequencies of 1341 and 1405 MHz, respectively. Data collected
before MJDs 53740 were folded every 16.7772\,s, and after that every
67.1088\,s. Off-source observations of a pulsed noise probe at
45$^{\circ}$ to the feed probes, but with otherwise identical
set-up, were taken at regular intervals for the purpose of
polarimetric calibration. Additionally, we analysed data for
PSR~J0437$-$4715 from the new Parkes Digital Filterbank (DFB)
system, a digital polyphase filterbank capable of 8-bit sampling. In
Table~\ref{tab:data} we present the details for all selected
datasets.

The data were then pre-processed with the \textsc{psrchive} software
package \citep{hvm04}. Specifically, for CPSR2 data we corrected the
2-bit digitisation artefact \citep{ja98} by applying the algorithm
in van Straten (in preparation), and removed 12.5\,\% of each edge
of the bandpass to avoid possible effects of aliasing and spectral
leakage. A full receiver model was determined and applied to perform
the polarisation calibration to the MB receiver data \citep{van04a},
as the receiver suffers from strong cross-coupling between the
orthogonal feeds. For the H-OH data from both back-ends we used the
common single-axis model instead, as the coupling was found to be an
order of magnitude weaker \citep[e.g.][]{mks+10}. The signals from
each polarisation were summed into total power (Stokes $I$), while
0.5\,MHz frequency channels were kept for later analysis purposes.
The template profiles used for the correlation coefficient
calculation and the cross-correlation procedure to estimate
$\sigma_{\rm rn}$ \citep{tay92}, were obtained independently from
the datasets shown in Table~\ref{tab:data}. All CPSR2 datasets have
passed through the test shown in \cite{lvk+11} to ensure no
significant 2-bit digitisation distortion is present. The test also
showed that the template matching produced the radiometer noise
uncertainty as expected by Eq.~(\ref{eq:sigma_rn}).

\section{Results}\label{sec:result}
\subsection{Measurement of shape constant}\label{ssec:SC-CC}
The measurement of the shape constant for individual integrations
gives an estimate of the profile stability, which here we have
carried out for the aforementioned five MSPs. For PSR~J0437$-$4715,
$\mathfrak{C}$ was determined using the the first 500 time dumps of
the MJD 53621 dataset. For PSRs~J1022+1001 and J1730$-$2304 we
integrated the time dumps into 1.0 and 1.8 minutes, separately, so
as to obtain sufficiently high SNR for calculation of $\mathfrak{C}$
and statistical analysis (see e.g. Fig.~\ref{fig:cctest}). The
$\rho$ is computed with respect to the on-pulse phase and the
root-mean-square (RMS) deviation of the noise is estimated based on
the off-pulse region, from which one can derive the value of
$\mathfrak{C}$ from Eq.~(\ref{eq:SCdef}). The expected values of
$\mathfrak{C}$ (denoted by $\mathfrak{C}_0$) are calculated directly
from the shape of the high SNR standards formed from previous
observations, which are also used in the calculations of $\rho$. The
errors in $\rho$ and the RMS of the baseline of the profile are
given by Eq.~(\ref{eq:sigmarho}) and by $\simeq{\rm
RMS}/\sqrt{2n_{\rm samp}}$, respectively.

\begin{figure}
\includegraphics[scale=0.3]{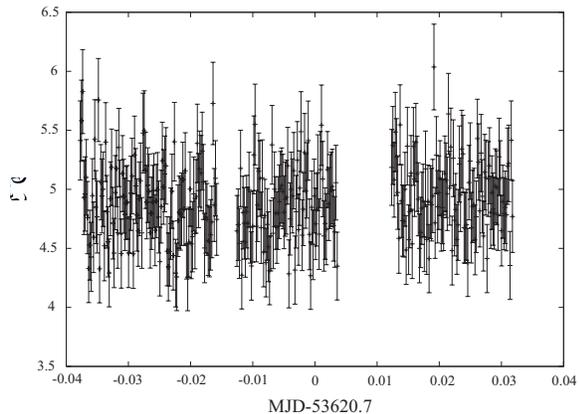}
\caption{Shape constant measurements for the first 500 integrations
of the MJD 53621 observation for PSR~J0437$-$4715. The time baseline
of these measurements is $\sim1.9$\,hours. \label{fig:tcc}}
\end{figure}

We present the shape constant measurement of PSR~J0437$-$4715 for
the MJD 53621 dataset in Fig.~\ref{fig:tcc} as an example, and show
the statistical result in Table~\ref{tab:SC stat}. It is clear that
for most sources, within the range of estimated error the measured
$\mathfrak{C}$ is in accordance with the analytical value.  The RMS
of the measured value also matches the mean error bar for each data
point. The deviation of the measured $\mathfrak{C}$ from the
expectation for PSR~1730$-$2304 may indicate an intrinsic level of
profile shape change, but could also be due to an insufficient
number of measurements. Note that, in this case, the statistics are
only based on 13 data points, while for each of the others more than
30 measurements were available.

It is worth noting that there has already been previous work
regarding the stability of PSR~J1022+1001 on both short and long
timescales, with inconclusive results \citep{kxc+99,rk03,hbo04}. On
timescales of roughly an hour, our results do not show profile
changes. However, the analysis was carried out using the whole
on-pulse phase, and may not be sensitive to variations occurring
only around the profile peaks. Any shape variations, if they exist,
are likely to affect mostly the peaks. A detailed analysis of the
statistics of the peak ratios will be published by Purver et al. (in
preparation).

\begin{figure}
\includegraphics[scale=0.3]{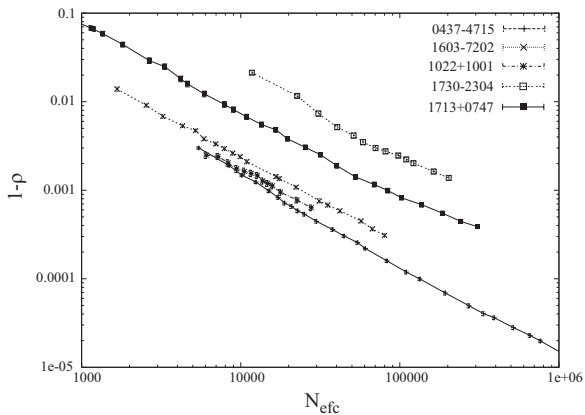}
\caption{$N_{\rm efc}-(1-\rho)$ plot for all five sources. The
effective pulse number is used to compensate for the varied weight
of each integration due to the difference in SNR. All curves show
expected linear behavior with fitted slopes to $-1$ in log-log
space. \label{fig:N-CC}}
\end{figure}

\begin{table}
\centering \caption{Statistical results of the profile shape
constant measurement for all five sources. The parameters $t_{\rm
int}$, $\mathfrak{C}_0$, $\overline{\mathfrak{C}}$, RMS,
$\overline{\sigma}$ and $k_{\rm NC}$ represent integration time of
the integrated profiles, shape constant calculated from the
standard, mean of measured $\mathfrak{C}$, RMS of measurements, mean
of individual measurement errors, and the fitted slopes of $N_{\rm
efc}$ versus $(1-\rho)$ curves in Fig.~\ref{fig:N-CC}, respectively.
}\label{tab:SC stat}
\begin{tabular}[c]{ccccccc} \\
\hline
Pulsar Name &$t_{\rm int}$ (s)&$\mathfrak{C}_0$ &$\overline{\mathfrak{C}}$ &RMS  &$\overline{\sigma}$ &$k_{\rm NC}$\\
\hline
J0437$-$4715  &16.8           &4.8              &4.9                       &0.32 &0.31                &-1.00(1)\\
J1022+1001    &67.1           &2.7              &2.6                       &0.22 &0.20                &-0.98(1)\\
J1603$-$7202  &16.8           &3.9              &4.0                       &0.33 &0.30                &-0.99(1)\\
J1713+0747    &16.8           &3.7              &3.8                       &0.23 &0.22                &-0.94(2)\\
J1730$-$2304  &117            &2.9              &3.2                       &0.24 &0.27                &-0.98(1)\\
\hline
\end{tabular}
\end{table}
To investigate the scaling relation indicated by
Eq.~(\ref{eq:SCdef}), for each source we perform incremental
integrations along the dataset, adding more pulse periods each time.
Our results are presented in Fig.~\ref{fig:N-CC} which shows the
relation between the weighted number of integrated pulses and
$\rho$. It is clear that all curves are linear in log-log space for
the $N_{\rm efc}-(1-\rho)$ relation as expected. The fitted slopes
all lie in the range of (0.94, 1.00), as given in the last column of
Table~\ref{tab:SC stat}. This result, together with the
$\mathfrak{C}$ statistics in Table~\ref{tab:SC stat}, indicates no
detectable profile shape variation along the integration.

\subsection{Fit of phase jitter}\label{ssec:fitf}
Estimation of the jitter parameter $f_J$ can be performed only on
the brightest source PSR~J0437$-$4715, as single pulse SNR for the
other sources is not high enough and radiometer noise is still
dominant in the timing residuals. Here we use all PSR~J0437$-$4715
datasets listed in Table~\ref{tab:data}, each of which contains over
a hundred individual integrations to yield a stable statistical
result. When performing the timing of the datasets, we used the
timing models derived by \citet{vbc+09}. TOAs and their
uncertainties were determined through cross-correlation with a
pre-formed standard, individually for each side of band. No EFAC or
EQUAD value were applied to change the measurement precision as done
commonly in timing analysis \citep[e.g.][]{hlj+11}.

\begin{figure}
\includegraphics[scale=0.35]{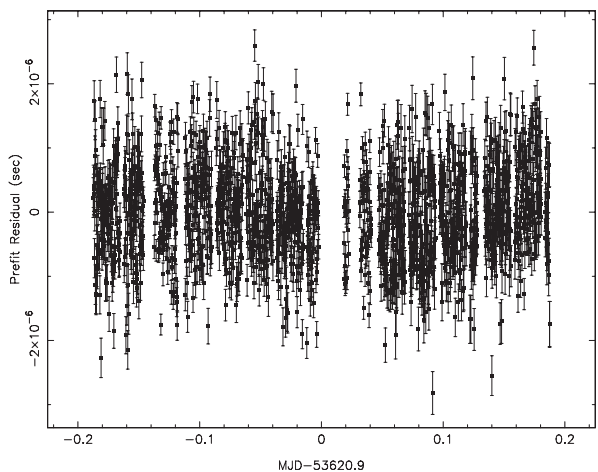}
\caption{The short-term timing solution of PSR~J0437$-$4715 over
$\sim$8 hours dataset on MJD 53621 at a central frequency of
1405\,MHz with 48\,MHz bandwidth. \label{fig:050907tim}}
\end{figure}

The timing solution of PSR~J0437$-$4715 shows a widely scattered
series of TOAs over a timescale of several hours, with a reduced
$\chi^{2}$ far larger than unity, which strongly indicates the
existence of extra uncertainty contributions besides radiometer
noise. Fig.~\ref{fig:050907tim} shows the timing residuals of the
MJD 53621 dataset for the 1405\,MHz band. The TOAs were produced by
summing over the whole bandwidth of effectively 48\,MHz, yielding
reduced $\chi^{2}$ values of 8.45 and 8.46 for the 1341\,MHz and the
1405\,MHz side band, respectively. Note that from
Eq.~(\ref{eq:sigma_scint}), the TOA fluctuation induced by unstable
profile broadening due to the ISM is approximately the scattering
timescale once the observational bandwidth is much narrower than the
scintillation frequency scale (thus $N_{\rm scint}\simeq1$). For
PSR~J0437$-$4715, as the scintillation bandwidth and pulse
broadening time are $\approx0.4$\,GHz and $\approx1$\,ns
\citep{cl02}, respectively, this uncertainty is negligible for the
datasets used in this paper.

Following the method mentioned in Section~\ref{ssec:jitter}, we
measure $f_J$ based on the shortest integrations from all
PSR~J0437$-$4715 datasets in Table~\ref{tab:data}, and the results
are summarised in Table~\ref{tab:fjmeasure}. It is clear that the
measurements from datasets collected by different receivers are
consistent with each other. This does not indicate significant
uncertainty contribution by polarimetric calibration error. Results
from two types of back-end also show consistency after correcting
for the bias induced by low-frequency noise of the CPSR2 data (see
the next paragraph for details). Additionally, the same result for
fits from both sides of the band suggests that the intensity of
phase jitter does not vary significantly on small frequency scales
(there is a 64\,MHz difference between the two bands) at observing
frequencies around 1.4\,GHz. The combination of all measurements for
both sides of the band achieves an estimated $f_J$ of
$0.067\pm0.002$. To study the statistics of the residuals, we
perform Kolmogorov-Smirnov (K-S) tests on each dataset, on TOAs
weighted by the modified uncertainties accounting for phase jitter.
The measured p-values \citep[e.g.][]{pftv86} are summarised in
Table~\ref{tab:fjmeasure}, not suggesting a significant deviation of
the weighted residuals from a Gaussian distribution. This
demonstrates the dominance of Gaussian noise in the residuals.

\begin{table}
\centering \caption{Results of jitter parameter $f_J$ measurements
and K-S tests of all PSR~J0437$-$4715 datasets from two 48-MHz
sub-bands. The parameters $f$ and $\mathcal {P}$ represent the
central frequency and the p-value of K-S test,
respectively.\label{tab:fjmeasure}}
\begin{tabular}[c]{cccc} \\
\hline
Dataset (MJD)  &$f$ (MHz) &$f_J$     &$\mathcal {P}$   \\
\hline
53576 &1341      &$0.067\pm0.004$ &0.89 \\
      &1450      &$0.067\pm0.004$ &0.93 \\
53621 &1341      &$0.067\pm0.004$ &0.98 \\
      &1405      &$0.066\pm0.004$ &0.62 \\
53964 &1341      &$0.066\pm0.004$ &0.88 \\
      &1405      &$0.065\pm0.004$ &0.92 \\
54095 &1341      &$0.069\pm0.006$ &0.90 \\
      &1405      &$0.073\pm0.006$ &0.92 \\
54222 &1341      &$0.069\pm0.007$ &0.52 \\
      &1405      &$0.067\pm0.007$ &0.77 \\
54226 &1341      &$0.068\pm0.005$ &0.20 \\
      &1405      &$0.066\pm0.006$ &0.64 \\
\hline
\end{tabular}
\end{table}

\begin{figure}
\includegraphics[scale=0.5]{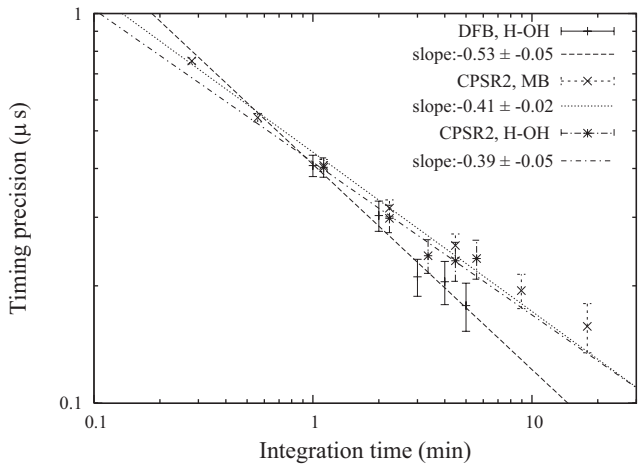}
\caption{Weighted timing residual versus integration time relation
for three combinations of instruments. The curve yielded by the DFB
data shows the expected power law with an index close to $-0.5$,
while the curves from CPSR2 data clearly deviate from this, which
implies additional uncertainty perhaps due to the low-bit
digitisation procedure. \label{fig:tintvrms}}
\end{figure}

To investigate the dependence of the result on the length of
individual integrations, we calculate the weighted RMS against
integration time $t_{\rm int}$ for multiple combinations of
instruments. In detail, MJD 53576 (CPSR2+MB), MJD 54095 (DFB+H-OH)
and MJD 54226 (CPSR2+H-OH) datasets are chosen to demonstrate
different hardware configurations, and the results are shown in
Fig.~\ref{fig:tintvrms}. Clearly, the curves yielded by CPSR2 data
collected by two receivers are consistent, again implying no
significant contribution of timing uncertainty by polarisation
calibration. The relation obtained from DFB data achieves a fitted
slope close to $-0.5$, supporting the idea from
Eq.~(\ref{eq:sigmas}) that timing residual scales as the square-root
of the number of integrated pulses once only the uncertainties by
radiometer noise and jitter are significant. The residuals from the
CPSR2 data coincide with those from the DFB in 1-min integrations,
and then saturate at the level of $\approx$150\,ns as the
integration time is extended. This implies that the difference may
be due to different digitisation procedures. The saturation
corresponds to additional self-correlated noise which contributes
$\approx$8\% bias in our $f_J$ measurement based on 1-min
integrations of CPSR2 data, which has been corrected for in the
results of Table~\ref{tab:fjmeasure}.

\begin{figure}
\includegraphics[scale=0.5,angle=-90]{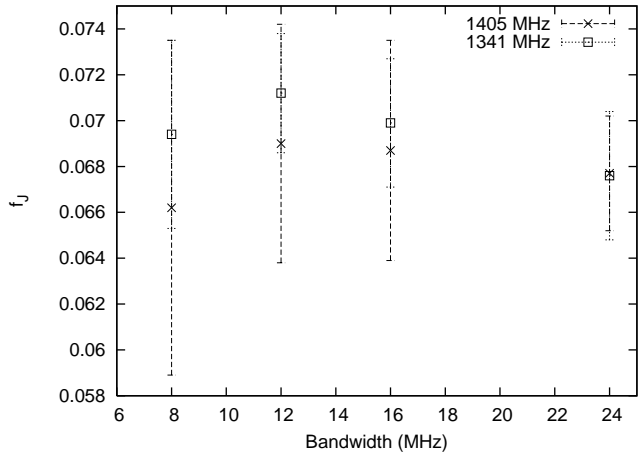}
\caption{Fitted amount of jitter against chosen bandwidth for both
sides of band from MJD 53621 dataset. No clear dependency between
these two parameters is visible. \label{fig:bwvf}}
\end{figure}

\begin{figure}
\includegraphics[scale=0.5,angle=-90]{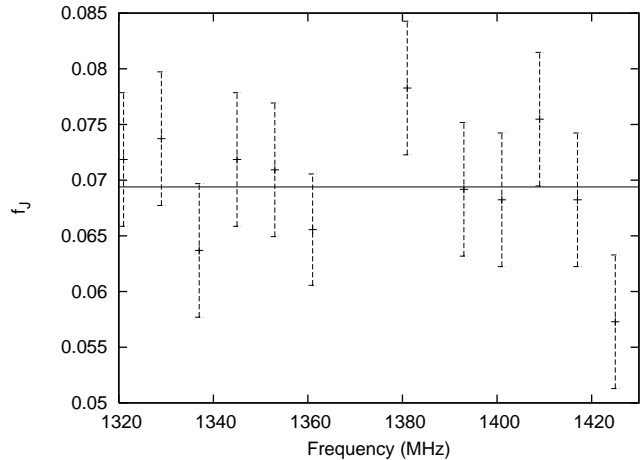}
\caption{Estimated jitter parameter based on an 8-MHz sub-bands from
the MJD 53621 dataset. The $f_J$ shows no clear trend of evolution
across a frequency range of $\sim100$\,MHz. \label{fig:freqvf}}
\end{figure}

The expression for $\sigma_{\rm J}$ indicates that the uncertainty
due to jitter is independent of the observational bandwidth and
central frequency. To illustrate this, the MJD 53621 dataset is
divided into two, three, four and six sub-bands each time. Then a
fit for the jitter parameter is carried out individually on each
sub-band and the results are combined incoherently to obtain an
estimated value of $f_J$ for a given bandwidth. The procedure is
carried out on both side-bands and the result is shown in
Fig.~\ref{fig:bwvf}. It is clear that at both central frequencies
jitter noise remains once the bandwidth is changed. The result
indicates that, on small frequency scales, pulses are jittered in
the same way, otherwise jitter noise would not remain the same after
summing the entire bandwidth. In Fig.~\ref{fig:freqvf} we plot the
fitted $f_J$ based on 8\,MHz bandwidth against the central frequency
for each sub-band. The group of $f_J$ values do not show a clear
dependence on the observational frequency, and yield average,
weighted RMS and reduced $\chi^2$ of 0.069, 0.006 and 1.1,
respectively. The mean correlation coefficient between residuals of
different sub-bands were calculated to be $\simeq0.51$, which
indicates a correlation among the time series and supports the idea
of equal jittering on small frequency scales.

\section{Conclusions and Discussions} \label{sec:conclu}
\subsection{Summary of the results} \label{ssec:sum}
In this paper, we investigate the issue of MSP profile stability
based on five pulsars in total. A shape constant associated with the
correlation coefficient is defined to quantify the stability. No
significant shape modulation of integrated profiles beyond the
measurement error is found for integration times from $10^1$ to
$10^2$\,s. For PSR~J0437$-$4715 we estimate the jitter parameter by
performing timing on short timescales and comparing the actual
timing residual with the amount expected from radiometer noise. The
fitted $f_J$ is found to be consistent on both sides of the bands
(64\,MHz difference in central frequency), and the combination of
several datasets results in an estimate of $0.067\pm0.002$. It is
also demonstrated that all the other potential sources of TOA
uncertainty, besides radiometer and jitter noise, do not strongly
influence the measurement. Additionally, we show that jitter noise
scales neither with bandwidth within a 50-MHz band nor with
frequency across a range of $\sim100$\,MHz at 1.4\,GHz, which
supports the idea that pulses are equally jittered on small
frequency scales. Such results, if still valid for wider frequency
range, would suggest that the jittering uncertainty cannot be
mitigated by extending the observing bandwidth \citep[also
see][]{ovh+11}.

\subsection{Future telescopes} \label{ssec:ftels}
\begin{figure}
\includegraphics[scale=0.33]{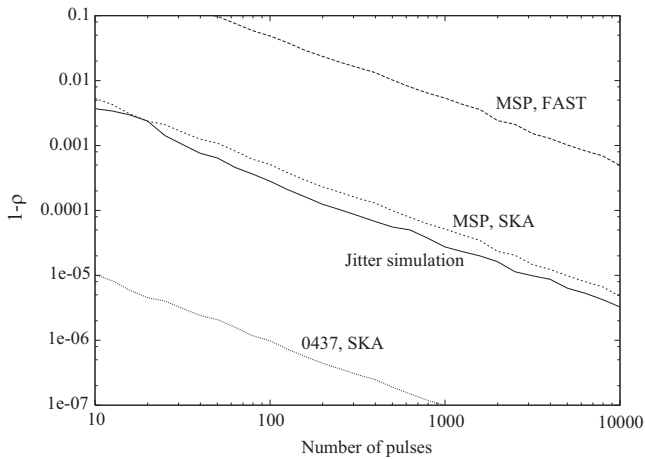}
\caption{$1-\rho$ versus $N$ curve prediction of PSR~J0437$-$4715
for different instruments and simulation. The pulses are created
with identical shape and a jitter parameter of $f_J=0.1$ is used to
simulate the Gaussian-distributed phase variation. \label{fig:NCCs}}
\end{figure}

Apart from bright single pulses and giant pulses
\citep[e.g.][]{cstt96,jak+98}, pulse jitter of the majority of MSP
pulses is still not detectable with currently available sensitivity.
However, with the next generation of radio telescopes, the shape
modulation of integrated profiles for some of the bright MSPs will
become visible. In Fig.~\ref{fig:NCCs}, based on the aforementioned
jitter model and assuming $f_J=0.1$, we perform a Monte Carlo
simulation to calculate the relation between the number of
integrated pulses and $1-\rho$ for the case of jitter only, and
compare the result with the curves from considering radiometer noise
only for a few instruments. We assume a 5-ms period, a 100-$\mu$s
pulse width, and a 5-mJy flux density at 1.4\,GHz for a sample MSP,
and 1.4\,GHz frequency with 300\,MHz bandwidth for observation. The
gains of FAST and SKA are assumed to be $20$\,$\rm m^2/K$ and
$100$\,$\rm m^2/K$, respectively \citep{nan06,sac+07}. For the
assumed jitter model the scaling also follows $1-\rho\propto N^{-1}$
as calculated in Appendix~\ref{app:scaling&err&j}. It is shown that
an SKA observation of an MSP of typical brightness, pulse jitter is
comparable to radiometer noise in influencing the
correlation-coefficient value. Note that in the applied jitter model
all single pulses are assumed to be identical, so the simulated
jitter curve can potentially move upwards once the shape modulation
is also accounted for. Future observations with the SKA of
PSR~J0437$-$4715 will be totally dominated by pulse jitter in shape
variation, so it will be ideal for single pulse study.

Once the shape variation of integrated profiles by pulse jitter
becomes significant, the current cross-correlation method for the
measurement of TOAs would fail in estimating the TOA uncertainty
correctly. Specifically, the model in the template-matching
procedure now needs to be of the form \citep{tay92}:
\begin{equation}
P(t)=A_0*S(t)+n(t)+j(t),
\end{equation}
where $P(t)$ is the observed profile, $S(t)$ is the template, and
$n(t)$ is the noise function. The parameter $j(t)$ is the
jitter-induced shape perturbation, which, referring to MSPs, is
mostly negligible compared with $n(t)$ for the current sensitivity.
If this shape difference becomes sufficiently large, the calculated
TOA precision would fail to follow the expected SNR scaling
\citep{lvk+11}. In this case, the shape modulation would need to be
modelled \citep[e.g., by principal component analysis,
see][]{cs10,ovh+11} and then a global determination of the unknown
parameters could be performed to properly estimate the TOA
\citep[e.g.][]{mldr11}.

\section*{Acknowledgments}
We thank J.~P.~W. Verbiest and B.~W.~Stappers for software support
and valuable discussions. We are also grateful to the anonymous
referee who provided constructive suggestions to improve the paper.
The Parkes telescope is part of the Australia Telescope which is
funded by the Commonwealth of Australia for operation as a National
Facility managed by CSIRO. KL is funded by a stipend from the
Max-Planck-Institute for Radio Astronomy.

\bibliographystyle{mnras}
\bibliography{journals,psrrefs,modrefs,crossrefs}

\appendix
\section{Correlation-coefficient Scaling for Additive Noise}\label{app:scaling&err&rn}
Assume the observed profile is a superposition of a normalised
template $\rm \bf p$ and Gaussian noises $\rm \bf n$, i.e.
$p_{i}=s_{i}+n_{i}$. The subscript $i$ goes from $1$ to $n$, the
number of the profile bins. We regard $\rm \bf p$ and $\rm \bf n$ as
$n$-dimensional vectors. Following Eq.~(\ref{eq:ccdef}), the
correlation coefficient between the perfect template $\rm \bf s$ and
observed profile $\rm \bf p$ is

\begin{equation}
  \rho=\frac{\displaystyle\sum_{i}c_{i}(c_{i}+n_{i})}{\displaystyle\sqrt{\sum_{i}c^{2}_{i}\sum_{i}(c_{i}+n_{i})^2
  }} \, ,
\end{equation}
where $c_{i}=s_{i}-\bar{s}$.

Assume that the noise $\rm \bf n $ is a multivariate Gaussian with
probability distribution $f({\rm \bf n})$ and covariance matrix $\rm
\bf C$, where $C_{ij}=\sigma_{\rm n}^2 \delta_{ij}$. The expectation
value of correlation coefficient $\langle \rho \rangle$ and its
second-order moment $\langle \rho^2 \rangle$ are
\begin{eqnarray}
 \langle \rho \rangle&=& \int \rho f({\rm \bf n}) \, d{\rm \bf
 n}\label{eq:avrrho}\\
 \langle \rho^2 \rangle&=& \int \rho^2 f({\rm \bf n}) \, d{\rm \bf n}
 \label{eq:avrrho2}
\end{eqnarray}
The variance for $\rho$ is then $\sigma_{\rho}^2= \langle \rho^2
\rangle- \langle \rho \rangle^2$. The
Eq.~(\ref{eq:avrrho})~and~(\ref{eq:avrrho2}) can be integrated by
transforming into hyper-spherical coordinates \citep{mw70}. Although
no analytical expression could be found, by using asymptotic
technique we derive the results for the case of a large sample
number which match both the low and high SNR cases as below:
\begin{eqnarray}\label{eq:CCexp}
\displaystyle1-\langle\rho\rangle&=&1-\left(\frac{1}{1+\chi}
\right)^{1/2} \\
&=&\displaystyle\left\{\begin{array}{cc} 1-(\frac{1}{\chi})^{1/2}+{\cal O}\left(\frac{1}{\chi^{3/2}}\right), &\chi \gg 1 \\
\displaystyle\frac{1}{2}\chi+{\cal O}\left(\chi^2\right), &\chi\ll 1
  \end{array}\right.
  \label{eq:onemrho}
  \end{eqnarray}

\begin{equation}
\label{eq:sigmarho}
  \sigma_{\rho}=\left \{\begin{array}{cc}
   \displaystyle \left[\frac{\chi}{n_{\rm samp}(1+\chi)}\right]^{1/2} + {\cal
O}\left(\frac{1}{\chi^2}\right), &\chi \gg 1  \\
   \displaystyle  \chi\left(\frac{n_{\rm samp}-1}{2n^{2}_{\rm samp}}\right)^{1/2} + {\cal
O}\left(\chi^2\right.), &\chi\ll 1
  \end{array}\right.
\end{equation}
where
\begin{equation}
  \chi=\frac{\displaystyle\sigma_{\rm
n}^2}{\displaystyle\overline{\sum_{i}c_{i}^2}},~~~~
\overline{\sum_{i}c_{i}^2}=\frac{\displaystyle\sum_{i}c_{i}^2}{n_{\rm
samp}}.
  \label{eq:defchi}
\end{equation}

\begin{figure}
\includegraphics[scale=0.5]{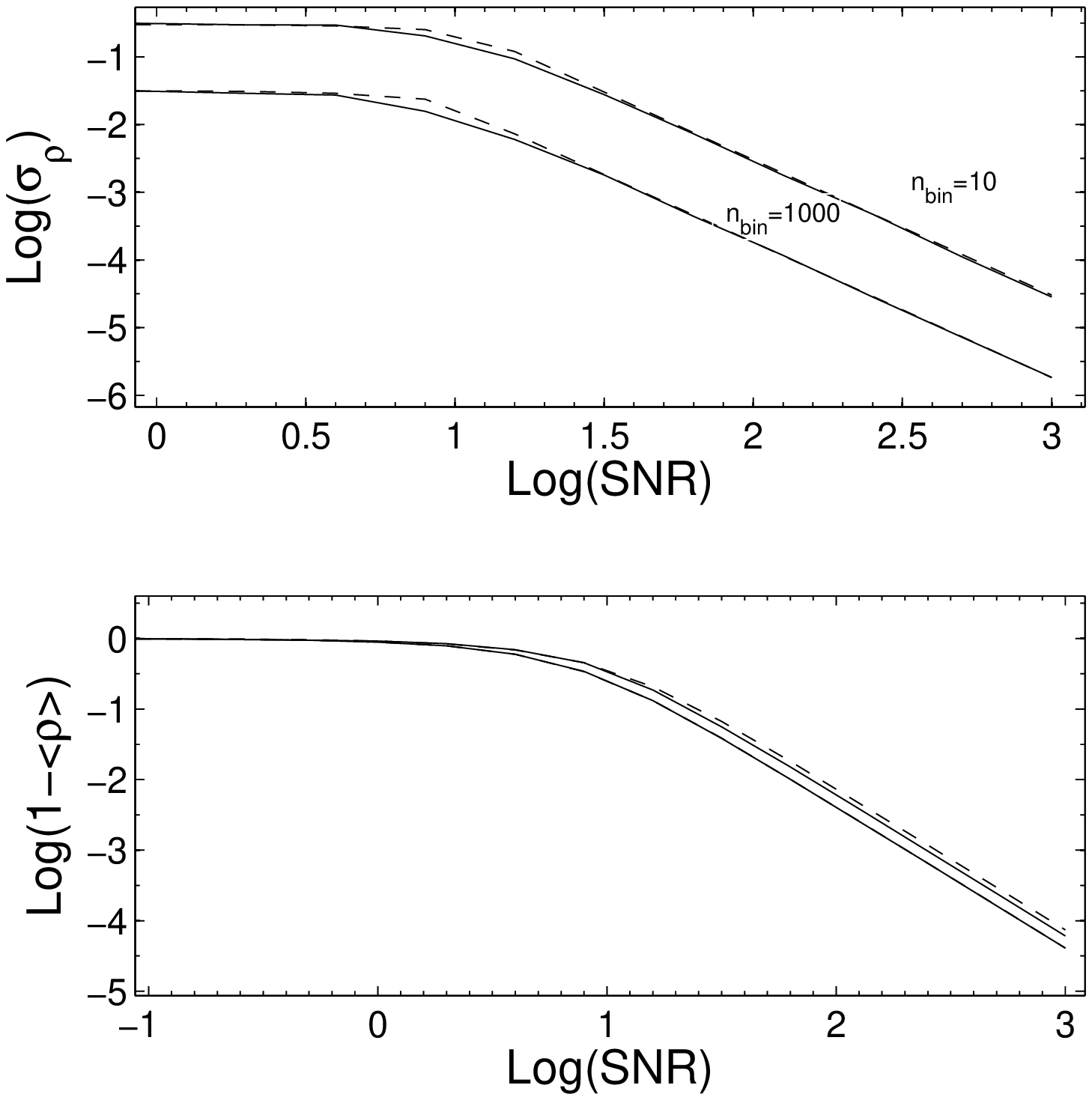}
\caption{Theoretical and simulated results presenting the relation
between SNR and correlation factors. In each case the dashed line
represents the analytic solution and the solid line stands for the
numerical result. The parameter $n_{\rm bin}$ is the number of
segments for the profiles. \label{fig:cctest}}
\end{figure}

In Fig.~\ref{fig:cctest} a set of Monte Carlo simulations are
performed so as to test the validity of the derivation. Here a
simple Gaussian template shape is assumed and profiles are created
by adding normal distributed noise onto the template. It can be seen
that for most SNR ranges the results from both approaches coincide
with each other, and for high SNR value, $1-\rho$ scales linearly
with the increase of signal.

\section{Correlation-coefficient Scaling for Jitter}\label{app:scaling&err&j}
In this section, we prove that the relation $1-\langle \rho\rangle
\propto 1/N$.  Assuming single pulses are of identical waveform
$p_{0}(\phi)$ and the effect of phase jitter is to introduce a
random phase to each single pulse, i.e. the $i$-th single pulse
takes a waveform of $p_{0}(\phi+\Delta \phi_{i})$, where the
$\Delta\phi_{i}$ is a random phase. The waveform of the template
profile $p(\phi)$ is defined by summing infinite number of single
pulses as
\begin{equation}
    p(\phi)=\lim_{N\to\infty}\frac{1}{N}\sum_{i=1}^{N}p_{0}(\phi+\Delta\phi_i)\,.
\end{equation}

Meanwhile, an integrated profile $p_{N}$ obtained from an
observational session is yielded by averaging $N$ single pulses
($N\gg1$) as
\begin{equation}
  p_{\rm N}(\phi)= \frac{1}{N}\sum_{i=1}^{N}p_{0}(\phi+\Delta\phi_{i})\,.
\end{equation}
The correlation coefficient between the template $p(\phi)$ and
$N$-averaged integrated pulse profile $p_{N}(\phi)$ is then
\begin{equation}
\rho=\frac{\displaystyle\int_0^1 {p(\phi)} p_{N}(\phi)\,d\phi}
{\left[\displaystyle{\int_0^1
 {p(\phi)}^2\, d\phi\int_0^1 p_{N}(\phi)^2\,d\phi}\right]^{\frac{1}{2}}}\,,
\label{eq:corefjit}
\end{equation}

The evaluation for the statistical expectation of correlation
coefficient is already presented in \cite{cs10}. In this appendix, a
slightly different approach is applied. Letting
\begin{equation}
  \delta(\phi)=p_{N}(\phi)-p(\phi)\,,
\end{equation}
we have
\begin{equation}
  \rho=\frac{\int_{0}^{1} p^2(\phi)+p(\phi) \delta(\phi)\,d\phi}{\left[
  \int_{0}^{1} p^2(\phi)\,d\phi  \int_{0}^{1} p^2(\phi)+2p(\phi)\delta(\phi)
  +\delta^2(\phi)\,d\phi\right]^{\frac{1}{2}}} \end{equation}
Since the number of pulse in observations is usually large, we
expect that the $p_{N}(\phi)$ is very close to $p(\phi)$, which
leads to $\delta(\phi)\ll p(\phi)$. In this case, one can show that
\begin{equation}
    \langle \rho\rangle=1+\frac{1}{2} \left\langle\left( \frac{\int_{0}^{1}
    p(\phi) \delta(\phi)
  \,d\phi}{ \int_{0}^{1} p^2(\phi)\,d\phi} \right)^2-\frac{\int_{0}^{1}
    \delta^2(\phi)\,d\phi}{\int_{0}^{1} p^2(\phi)\,d\phi}\right\rangle+{\cal
  O}\left(\frac{\delta^4}{p_0^4}\right)\,, \end{equation}
which is equivalent to
\begin{equation}
    \langle 1-\rho\rangle \simeq\frac{1}{2} \left\langle \frac{\int_{0}^{1}
  p_{N}^2(\phi)\,d\phi}{\int_{0}^{1} p^2(\phi)\,d\phi}- \left(
    \frac{\int_{0}^{1} p(\phi) p_{N}(\phi) \,d\phi}{ \int_{0}^{1}
        p^2(\phi)\,d\phi} \right)^2\right\rangle\,. \end{equation}

To understand how the $1-\langle \rho \rangle$ depends on the number
of pulses $N$, we have to expand $p_{N}(\phi)$ in the above
equations. One can then have \citep{cs10}
\begin{eqnarray}
    \left\langle\int_{0}^{1} p_{N}^2(\phi) \,d\phi\right\rangle&=&\frac{N^2-N}{N^2}\int_{0}^{1}
    p^2(\phi)\,d\phi \nonumber \\
    &&+\frac{1}{N}\int_{0}^{1}
    \langle p_{0}^2(\phi+\Delta\phi)\rangle\,d\phi\,,
\end{eqnarray}
and
    \begin{eqnarray}
    \left\langle\left(\int_{0}^{1} p(\phi) p_{N}(\phi) \,d\phi\right  )^2
    \right\rangle = \frac{N^2-N}{N^2}\int_{0}^{1} p^2(\phi)\, d\phi \nonumber \\
  +
    \frac{1}{N} \int_{0}^2\,d\phi\int_{0}^1\,d\phi' p(\phi)p(\phi')\langle
    p_0(\phi+\Delta\phi) p_0(\phi'+\Delta\phi) \rangle\,.
\end{eqnarray}
Here we assume that the phase jitter is independent, i.e. $\langle
\Delta \phi_{i} \Delta \phi_{j}\rangle=0$, if $i\neq j$. Thus

\begin{equation}
  \langle 1-\rho \rangle= \frac{1}{2N}\left[ S-K\right] \,
  \label{eq:eqsk}
\end{equation}
where
\begin{eqnarray}
    S&=&\frac{ \int_{0}^{1} \langle p_{0}^2(\phi+\Delta\phi) \rangle\, d\phi}{
    \int_{0}^{1} p^2(\phi)\,d\phi}\,, \\
    K&=&\frac{\int_{0}^2\,d\phi\int_{0}^1\,d\phi' p(\phi)p(\phi')\langle
    p_0(\phi+\Delta\phi) p_0(\phi'+\Delta\phi) \rangle}{ \left(\int_{0}^{1}
    p^2(\phi)\,d\phi\right)^2}\,. \nonumber \\
    &&
\end{eqnarray}
Because neither $S$ nor $K$ is $N$-dependant, the
Eq.~(\ref{eq:eqsk}) clearly show the scaling relation that
$1-\langle \rho \rangle\propto 1/N$.

\begin{figure}
\includegraphics[scale=0.4]{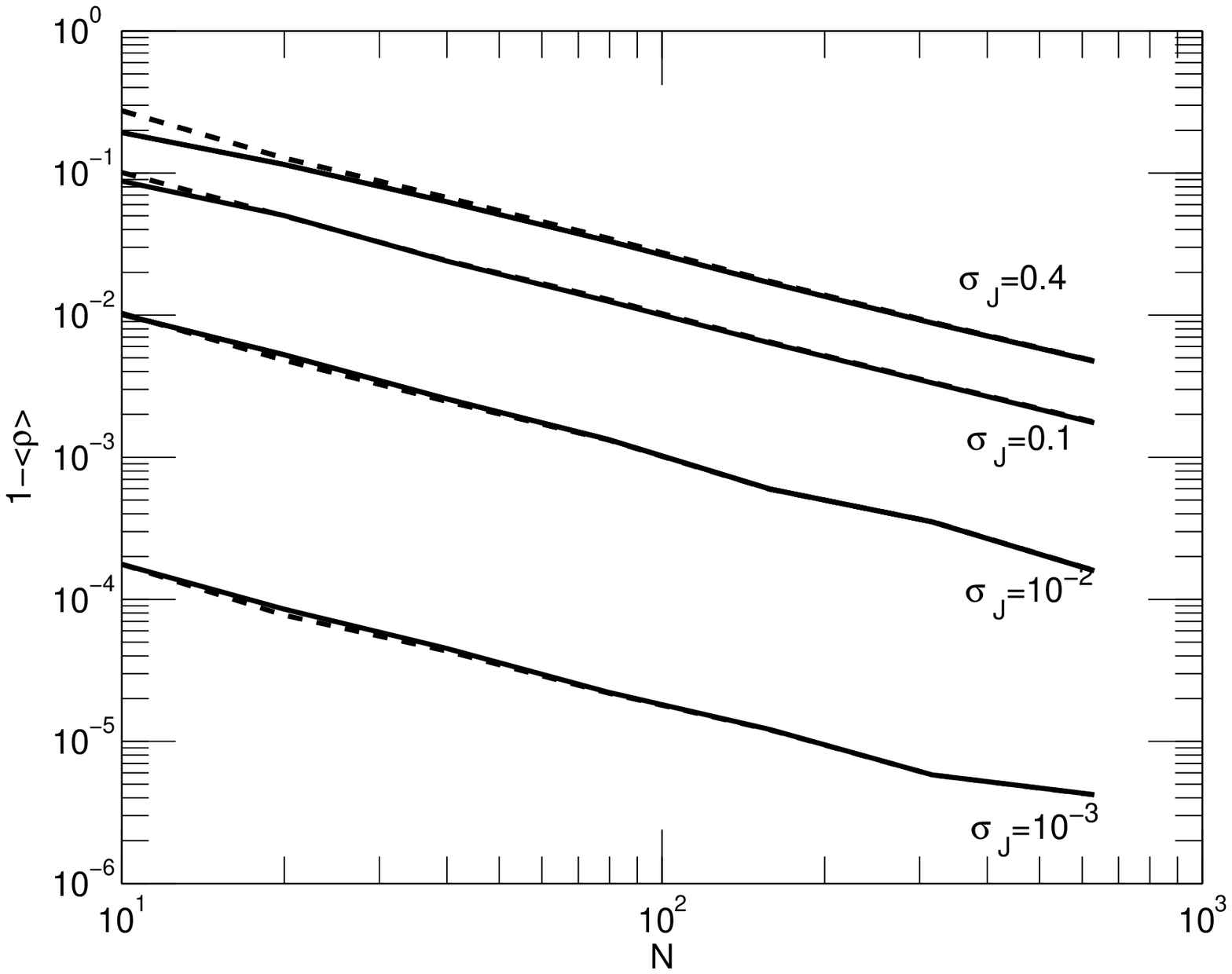}
\caption{The $1- \langle\rho\rangle$ as function of $N$ and
$\sigma_{\rm J}$, where $\sigma^2_{\rm J}$ is the variance of the
phase jitter probability density function. The dashed lines are from
analytical calculation, the Eq.~(\ref{eq:eqsk}), while the solid
lines are from direct Monte Carlo simulations. \label{fig:comp} }
\end{figure}

We also further compare the result of $1-\langle \rho \rangle$ form
Monte Carlo simulations and from the Eq.~(\ref{eq:eqsk}) in the
Fig.~\ref{fig:comp}. Clearly, the numerical simulation and
analytical calculation match each other at larger $N$ limit as we
expected. 
\end{document}